\documentclass{cjaa}           

\usepackage{graphicx,times}                   
\input{epsf.sty}                        
\input{psfig.sty}                       
\begin{document}
   \title{Mid-to-Far Infrared Spectral Energy Distribution of Galaxies in {\it Spitzer} First Look Survey Field
$^*$ \footnotetext{\small $*$ Supported by the National Natural
Science Foundation of China.} }

\volnopage{Vol.\ 7 (2007), No.\ 2,~ 000--000}
   \setcounter{page}{1}
   \author{Xiao-Qing Wen
      \inst{1, 2}
   \and Hong Wu
      \inst{2}
   \and Chen Cao
      \inst{2}
   \and Xiao-Yang Xia
      \inst{1, 2}
      }
   \institute{Department of Physics, Tianjin Normal University, Tianjin 300074\\
        \and
             National Astronomical Observatories, Chinese Academy of Sciences, Beijing
             100012;
             {\it hwu@bao.ac.cn}\\
\vs\no
 {\small Received 2006 April 18; accepted 2006
 June 7}
 }

\abstract{ We
made model fitting to the mid-to-far infrared
spectral energy distributions (SEDs) for different categories of
galaxies in the main extragalactic field of the {\it Spitzer}
First Look Survey  with the aid of spectroscopic information from
the Sloan Digital Sky Survey. We find that the mid-to-far infrared
SEDs of HII galaxies, mixture type galaxies and LINERs can be well
fitted by
the
one-parameter ($\alpha$) dust model of Dale et al.
plus
the 13 Gyr dust-free elliptical galaxy model. The statistics
of
$\alpha$ values indicates that all these galaxies tend to be
quiescent, although
the HII galaxies are relatively more active than
the LINERs. The mid-infrared SEDs of absorption galaxies are well
fitted simply by the 13 Gyr dust-free elliptical galaxy template,
and the near-to-mid infrared  SEDs of QSOs can be represented by
AGN NGC~5506.  }

   \authorrunning{X. Q. Wen et al.}            
   \titlerunning{Mid-to-Far Infrared SEDs of Spitzer Galaxies}  

   \maketitle

\section{Introduction}           
\label{sect:intro}

To understand the star formation process, it is essential to
investigate the infrared spectral energy distributions (SEDs) of
galaxies (Sajina et al. 2006). The mid-to-far infrared emission
powered by either stars or AGNs is generated by
dust
reprocessing and contributes
from a few
percent to more than 95 percent of
the total energy output (Xilouris et al. 2004; Soifer et al. 1984;
Dale et al. 2005). Infrared emission is much important at higher
redshift where star formation is more
frequent. The infrared SEDs can help to understand the cosmic infrared background (Dale et al.\ 2001) and provide information about the galaxy evolution as well. However, the detection limit in infrared bands prevents us
from exploiting the infrared SEDs of normal galaxies at higher
redshifts, hence the infrared SEDs of normal galaxies can only be
obtained
for
the local Universe.

Rowan-Robinson et al. (2004) presented three to four bands
rest-frame infrared SEDs of several tens of galaxies from the
final band-merged European Large-Area ISO Survey (ELAIS) catalog
at 6.7, 15, 90 and 175\,$\mu$m. All the infrared SEDs can be
fitted by cirrus, starburst (M82), ultraluminous infrared galaxy
(Arp220) and active galactic nuclei (AGN) dust torus models.
Dennefeld et al. (2005) identified 28 170\,$\mu$m sources from
FIRBACK N1 ISO survey and obtained the SEDs from the mid-infrared
to submilimeter bands. They found that the M82 template failed to
fit FIRBACK sources, instead, most FIRBACK sources could be well
fitted by the normal galaxy template (cold galaxies) from Lagache
et al.\ (2003). Some of sources can be well fitted by
the one parameter dust model (Dale et al. 2001) with $\alpha$
$>$ 1.6,
indicating
dominance of the quiescent component.
They concluded that most of the identified FIRBACK sources are quiescent star-forming galaxies, rather than
starbursters.

The Infrared Array Camera (IRAC: Fazio et al. 2004) and the
Multiband Imaging Photometer of {\it Spitzer} (MIPS: Rieke et al.
2004) instruments are designed for covering the range from the
mid-infrared to far-infrared bands and can provide the highest
sensitivity and resolution in mid-infrared (about 2$\arcsec$ in
IRAC images and 6$\arcsec$ in MIPS 24\,$\mu$m image). The
successful launch of {\it Spitzer Space Telescope } (Werner et al.
2004) provides a new opportunity to explore the mid-to-far
infrared SEDs of
a larger number of galaxies. With the new data from
{\it Spitzer} Legacy Program SWIRE (Lonsdale et al. 2003),
Rowan-Robinson et al. (2005) modeled the SEDs of selected ELAIS
sources in N1 field with cirrus, M82, Arp220 and AGN dust torus
models. Dale et al. (2005) provided 1--850\,$\mu$m SEDs of 75
galaxies from the {\it Spitzer} Infrared Nearby Galaxies Survey
(SINGS, Kennicutt et al. 2003), from the data of {\it Spitzer},
2MASS, ISO, IRAS and SCUBA. They successfully fitted the SEDs with
the one parameter dust model (Dale et al. 2001).

Although the infrared SEDs of many galaxies can be well fitted by
Rowan-Robinson et al. (2004,2005) and Lagache et al. (2003), their
models are based on the SEDs of
representative objects
such as M82 and Arp 220. Thus, these models could not provide further
quantitative analysis. Therefore, it is necessary to construct a
more quantitative model. Dale et al. (2001) introduced a
one-parameter dust model which can describe the infrared SEDs in a
simple
and a quantitative way. Dale \& Helou (2002) extended the
empirical calibration to long wavelength by new far-infrared and
submilimeter data. In this model,
a parameter $\alpha$,
representing the activity of the galaxy,
characterizes the shape of
the infrared SEDs.
To quantitatively explore the infrared properties of
galaxies with different optical spectral types, we adopt the dust
radiation model of Dale et al.\ (2001) in this work.

The structure of present work is as follows. In Section 2 we
describe the sample, data reduction and optical spectral
classification. The mid-to-far infrared SEDs of sample galaxies
and corresponding fitting are given in Section 3. Results and
discussion are presented in Section 4 and conclusions are drawn in
Section 5. Throughout this paper we adopt a cosmology with a
matter density parameter $\Omega m$=0.3, a cosmological constant
$\Omega A$=0.7 and a Hubble constant of $H 0$=70~km~s$^{-1}
$Mpc$^{-1}$.

\section{The Sample and Spectral Classification}
\label{sect:Sams}
\subsection{The Sample}

The galaxy sample was taken from the extragalactic component of
the main field of the {\it Spitzer Space Telescope} (Werner et al.
2004) First Look Survey (xFLS). Observations covering the survey
areas were made with both IRAC (Lacy et al. 2005) and MIPS (Frayer et al. 2006). The overlap area of xFLS imaged by both IRAC and
MIPS is about 3.7 degree$^2$. The BCD (Basic Calibrated Data)
images of IRAC four bands were  obtained from {\it Spitzer}
Sciences Center, which include flat-field corrections, dark
subtraction, linearity and flux calibrations (Fazio et al. 2004).
The IRAC images (in all four IRAC bands: 3.6, 4.5, 5.8 and
8.0\,$\mu$m) were
mosaicked from the BCD images after pointing
refinement, distortion correction and cosmic-ray removal with the
final pixel scale of 0.6 $\arcsec$ as described by Huang et al.
(2004) and Wu et al. (2005); whilst the MIPS 24\,$\mu$m images
were mosaicked in
a similar way with the final pixel scale of 1.22$''$. Matching the sources detected by SExtractor (Bertin \& Arnouts 1996) in five bands (IRAC four bands and MIPS 24\,$\mu$m band ) with the 2MASS sources, we derived
an astrometric uncertainty of $\sim$0.1$\arcsec$.

As this area has also been covered by the Sloan Digital Sky Survey
(SDSS: Stoughton et al.\ 2002), the xFLS sources were then matched
with all the galaxies with SDSS spectroscopic observations with a
radius of 2 $\arcsec$. Finally, a sample of 315 sources were
obtained, which includes 270 galaxies and 45 QSOs
(including Seyfert 1s). The ''AUTO'' magnitudes (from SExtractor) were adopted to
represent the total magnitudes of all the sample galaxies in IRAC
four bands and MIPS 24\,$\mu$m band. The IRAC and MIPS 24\,$\mu$m
bands have flux calibration accuracies better than 10\% (Fazio et
al. 2004; Rieke et al.\ 2004). The matched MIPS 70\,$\mu$m and
160\,$\mu$m fluxes (with a radius of 5$\arcsec$) of all these
sources were taken from Frayer et al.'s (2006) catalog and have a
signal-to-noise ratio better than 7. Here, 79 of 270 galaxies and 8
of 45 QSOs have obtained 70\,$\mu$m fluxes; whilst 40 galaxies and
only one QSOs have obtained
160\,$\mu$m fluxes. The absolute flux uncertainties of these two bands are 15\% and 25\%, respectively.

\subsection{Spectral Classification}

Before the spectral classification we carried out the extinction
correction, including the foreground extinction of the Milky Way
and the intrinsic extinction of the galaxy itself. The Milky Way
extinction is corrected by the parameterized curve of Calzetti et
al.\ (2000) and the intrinsic extinction is derived from the
Balmer decrement $F {\rm H\alpha}/F {\rm H\beta}$ (Calzetti 2001).
Here, we adopted
intrinsic ratios $I(\rm H\alpha)/I(\rm H\beta)=2.85$ for HII galaxies and
$I(\rm H\alpha)/I(\rm H\beta)=3.1$ for AGNs (Veilleux \& Osterbrock 1987; Wu et al.\ 1998). The optical spectra were obtained with 3\arcsec aperture
fibers (Stoughton et al. 2002) and the corresponding fluxes of
the emission lines were taken from the SDSS catalog of measured
line fluxes (Tremonti et al.\ 2004, version 5.0$\ $4).

The spectral classification of galaxies is based on the
traditional line-diagnostic diagram [NII]/H$\alpha$ versus
[OIII]/H$\beta$ (Veilleux \& Osterbrock 1987), which is
parameterized by Kauffmann et al. (2003). To derive
a
reliable spectral classification, we only dealt with
galaxies with relative strong H$\alpha$ emission (EQW(H$\alpha) \le -1.5$, here,
the negative value represents the emission line) and enough
signal-to-noise ($F {\rm H\beta} \ge 3\sigma$). If a source
position is
close
to the boundary between HII galaxies and LINERs
on the diagnostic diagram with the separation less than one sigma,
we classified it as
a mixture type
galaxy (see  Wu et al. 1998). The diagnostic diagram of the sample galaxies
is
shown in Figure~1.  The galaxies with H$\alpha$ absorption line were
treated as absorption galaxies.  Most of them are early type
galaxies. Finally we obtained 69 HII galaxies, 28 LINERs, 2
Seyfert 2s, 33 mixture types, 27 absorption galaxies and 45 QSOs.
There are 111 unclassified galaxies,
because
of their relatively
weak
H$\alpha$ emission or lower signal-to-noise ratio. It should be
noted that the Seyfert 2 galaxy SDSS J171544.03+600835.3 marked in
Figure~1 is a spectral double-peak galaxy with
a velocity
separation of about 300~km~s$^{-1}$. Since
its fluxes
are not
given in
Tremonti et al.'s (2004) catalog,
we measured the emission lines needed by {\it Splot} task in IRAF. During the measurement, we treat the double peak lines as one. In Table~1 we give the
number of sources detected by either IRAC or MIPS bands for
different spectral types of sources in xFLS.

\begin{figure}
\centering
\includegraphics[width=100mm]{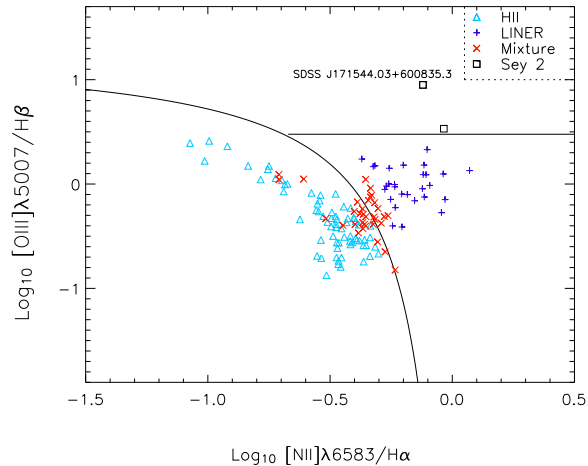}

\caption{\baselineskip 3.6mm
 Diagnostic diagram of spectral classification
[NII]$\lambda$6583/H$\alpha$ versus [OIII]$\lambda$5007/H$\beta$
(Veilleux \& Osterbrock 1987). The solid curve is the
parameterized curve by Kauffmann et al. (2003). Triangles are HII
galaxies, plus signs are LINERs,  crosses are mixture types, and
open boxes are Seyfert 2 galaxies.} \label{fig1}
\end{figure}

\subsection{The Mid-Infrared Color-Color Diagram}

Figure 2 shows the mid-infrared color-color diagram ([3.6]--[4.5]
vs. [3.6]--[8.0]) for all the classified galaxies. The magnitudes
used here are in
the AB magnitude system (Oke \& Gunn 1983). It is
clear from Figure~2  that all
the QSOs are well
separated from
the other galaxies. This can be explained
by the emission of QSOs at 3.6\,$\mu$m and 4.5\,$\mu$m
being dominated by
the power-law continuum from the dust torus of AGN, and the emission of normal galaxies at these two bands
being dominated by the decreasing stellar continuum after the peak in near-infrared. Thus, the [3.6]--[4.5] colors of QSOs are much redder than those of normal galaxies. In fact, [3.6]--[4.5] color is one of the best
parameters to select QSOs from galaxies (Hatziminaoglou et al.
2005). On the other hand, as the 8\,$\mu$m band covers the
strongest 7.7\,$\mu$m PAH (polycyclic aromatic hydrocarbons)
emission, the [3.6]--[8.0] color can roughly characterize the
ratio of PAH emission to old stellar continuum. The stronger the
PAH emission, the redder the [3.6]--[8.0] color. It is obvious
from Figure~2 that the absorption galaxies are separated from all
other galaxies
in
being
located at the lower-left part of Figure~2. In
contrast, all the other galaxies, such as HII galaxies, LINERs,
mixture type galaxies and Seyfert 2s,
are
spread
over a larger range of
[3.6]-[8.0] color
in upper part of Figure~2, indicating
that they could have contribution from PAH emission more or less
contrary to
the early type galaxies.

\begin{figure}
\centering
\includegraphics[width=100mm]{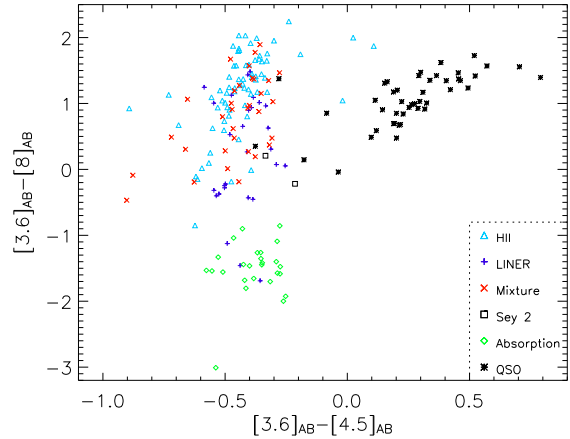}
\vspace{-6mm}

\caption{\baselineskip 3.6mm  Mid-infrared color-color diagram
([3.6]--[4.5] vs. [3.6]--[8]) of the sample galaxies. Triangles
are HII galaxies, plus signs are LINERs, crosses are mixture
types, squares are Seyfert 2s, diamonds are absorption galaxies,
and stars are QSOs. We adopt AB magnitude system (Oke \& Gunn
1983). QSOs are well separated from other galaxies and the
absorption galaxies are located at the left-bottom.}
   \label{fig2}
   \end{figure}

Detailed properties of all the sample galaxies are analyzed
from their SEDs
in the following sections.

\section{Spectral Energy Distribution}
\subsection{SED Templates}

To understand the SEDs of the different types of galaxies in this
sample, several model templates
based
on different radiation mechanisms are used. (1) Template of NGC~5506 is adopted as the template of QSOs. (2) The SED of 13 Gyr dust-free elliptical
galaxy, which is obtained from GRASIL code (Silva et al. 1998), is
adopted as the template of the absorption galaxies. (3) To
describe the emission line galaxies, such as HII galaxies,
LINERs, mixture types galaxies and even Seyfert 2s, a pure dust
model
plus an underlying stellar component are needed. Since
the old stellar population dominates the mid-infrared stellar
continuum emission in most normal galaxies, an SED of 13 Gyr
dust-free elliptical galaxy is also adopted as the stellar
component of emission line galaxy. Meanwhile, the
one-parameter
dust model (Dale et al. 2001) is used,
with the
one parameter
$\alpha$
defined essentially by the far-infrared color.
One can construct a sequence of dust model templates with different power index $\alpha$. The parameter $\alpha$
characterizes the shape of mid-to-far infrared  SEDs and relates with the activity of
the galaxy (Dale et al. 2001).

\subsection{$\chi^2$  Fitting}

We performed the model fitting to
the observed mid-to-far infrared SED for each galaxy. To quantitatively analyze the contributions of various radiation components, we assumed that
radiations of galaxies in
these bands come from
either one or
a combination of these radiations. With the aid of
well built SED template for each radiation component, a standard
$\chi^2$ minimum can be carried out to obtain the best fitting.
That is,
\begin{displaymath}
\chi^2=\sum {i=1}^{n {\rm band}} \frac{[f {i}({\rm obs})- c
f {{\rm temp}, i}(p)]^2}{\sigma {i}^2} \label{eq:Lebseque4}.
\end{displaymath}
where $f {i}(\rm obs)$ and $f {{\rm temp}, i}(p)$ are the observed
and model template fluxes at {\it i}-th band, {\it c} is
a scale
parameter
related
with the observed flux, {\it p} is the
model parameter, which is the $\alpha$ parameter in Dale et al.'s
dust model, $n {\rm band}$ is the number of bands used, and
$\sigma i$ is the observed error, which includes both random error
and calibration error. The minimum of $\chi^2$ of a set of models
was
taken as the best fit of an observed SED.

\subsection{Modeling the SEDs of Galaxies}

The model templates described above
cover the main components of radiations in mid-to-far infrared bands, i.e., radiations from old stellar population, dust (including PAH) or dust torus of AGN.
Because our sample includes different types of galaxies, we will
describe their SEDs, fitting processes and results of each type of
galaxies, separately.

\begin{table}
\centering
\begin{minipage}{100mm}
\caption[]{Number of Different Types of Galaxies Detected in the
IRAC and MIPS Bands}\end{minipage}\vs

\fns
\tabcolsep 3mm
\begin{tabular}{lccccc}
  \hline\noalign{\smallskip}
        &  $N {\rm total}$ &  $N {\rm IRAC}$ & $N {24\,\rm \mu m}$  &
        $N {\rm 70\,\mu m}$ &   $N {\rm 160\,\mu m}$ \\
  \hline\noalign{\smallskip}
HII     &  69  &  69   &     61       &        46     &     20         \\
LINER   &  28  &  28   &     24       &        13     &     5        \\
Mixture &  33  &  33   &     30       &        16     &     9 \\
Seyfert 2   &  2   &  2    &     2        &        1     &     0   \\
Absorption     &  27  &  27   &     3        &        0     &     0 \\
Unclassified   &  111 &  111  &     43       &        3      &     6      \\
QSO     &  45  &  45   &     43       &        8      &     1     \\
  \noalign{\smallskip}\hline
  \end{tabular}
\end{table}

Almost all of the HII galaxies, LINERs, mixture type galaxies and
Seyfert 2s have been detected in 24\,$\mu$m band (see Table 1).
In
contrast,
because of the 7$\sigma$ limit of
the MIPS catalog (Frayer et al. 2006),
only about
half
of the target galaxies
have
70\,$\mu$m fluxes (46 of 69 HII galaxies, 13 of 28 LINERs, 16 of
33 mixture types and 1 of 2 Seyfert 2s) and only a few galaxies
have
160\,$\mu$m fluxes (20 HII galaxies, 5 LINERs, 9 mixture
types). These
detection
statistics
are shown in Table~1. We obtained the individual rest-frame SEDs and show some
exemplar
HII galaxies, LINERs, mixture types and Seyfert 2 galaxies
in Figure~3.

\begin{figure}
\centering

\includegraphics[width=120mm]{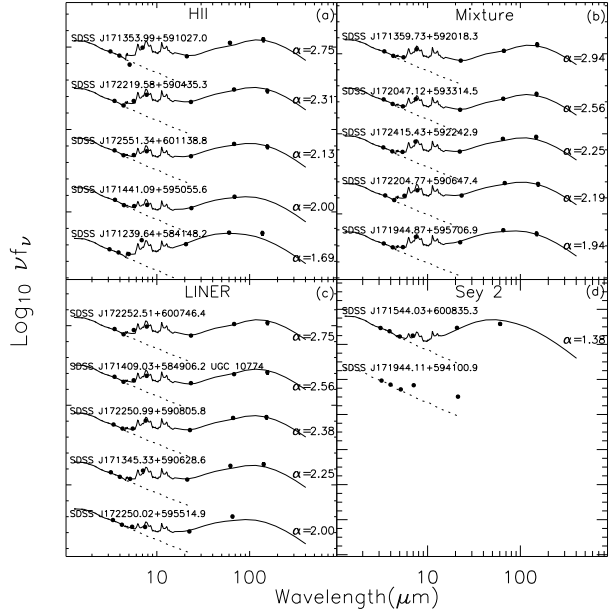}

\caption{\baselineskip 3.7mm Exemplar SEDs of HII galaxies,
LINERs, mixture type galaxies and Seyfert 2s. The dotted lines are
the 13 Gyr dust-free elliptical galaxy synthesized from GRASIL
code (Silva et al. 1998). The solid lines are the best fitted
composite SEDs of Dale et al.'s dust model (2001) plus a 13 Gyr
dust-free elliptical galaxy model. All the HII galaxies, LINERs,
and mixture type galaxies are well fitted by this model. The
double-peak Seyfert 2 galaxy SDSS J171544.03+600835.3 can be also
roughly fitted. } \label{fig3}
\end{figure}

During the model fitting we simply assumed that all the
3.6\,$\mu$m and 4.5\,$\mu$m fluxes come from the underlying
stellar component. The first step was to obtain the model
underlying stellar component based on the $\chi^2$ fitting to
3.6\,$\mu$m and 4.5\,$\mu$m fluxes. Then the $\chi^2$ method was
used again to find the best dust model to the fluxes with the
stellar component subtracted in the other bands. Note that we only
did the fitting to galaxies with at least one far-infrared flux
(either MIPS 70\,$\mu$m or 160\,$\mu$m band), because $\alpha$ is
used to characterize the shape of mid-to-far infrared SEDs, rather
than that of only mid-infrared SEDs. The best fitting models are
plotted as solid lines and the model stellar components are
plotted as dotted lines, in Figure~3.

We can see from Figure 3 that the model generally represents well
the mid-infrared PAH features and far-infrared dust continuum for
the HII galaxies, LINERs and mixture types, except some deviation
appearing at the long wavelength (160\,$\mu$m)
in a few galaxies.
This deviation can be explained by the lower weight (the larger
error) of 160\,$\mu$m during the fitting. The SED of spectral
double-peak Seyfert 2 galaxy, SDSS J171544.03+600835.3, can also
be roughly fitted but with some deviation in
the 70\,$\mu$m band. The
low $\alpha$ value of 1.38 indicates that this galaxy is very
active, and possibly could be explained by its active nuclei.

Almost all of the absorption galaxies escaped detection by MIPS:
only three of them were detected in the 24\,$\mu$m band (see Table
1). The rest-frame SEDs of all the absorption galaxies are plotted
in Figure~4  and scaled by the rest-frame 3.6\,$\mu$m flux.
Generally, their mid-infrared SEDs present small scatter and can
be well fitted by a 13 Gyr dust-free elliptical galaxy (solid line
in Fig.\,4) synthesized from GRASIL (Silva et al. 1998). This
indicates that these galaxies are dominated by the old stellar
population in mid-infrared band with little dust contribution. The
absorption galaxies present a much larger scatter in the longer
wavelength ($\lambda$ $>$3\,$\mu$m) than in the short wavelength
($\lambda$ $<$3\,$\mu$m). This could be the result of weak dust
emission, such as diffuse PAH emission.

\begin{figure}
\centering
\includegraphics[width=80mm]{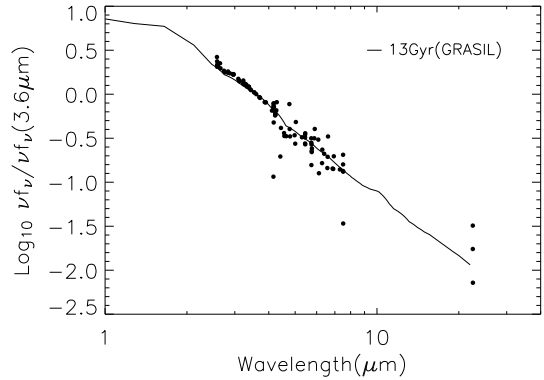}

\caption{\baselineskip 3.6mm  SEDs of the absorption galaxies with
fluxes scaled at rest-frame 3.6\,$\mu$m. The line is the 13 Gyr
dust-free elliptical galaxy model from GRASIL. Such an old stellar
population model can fit the absorption galaxies very well.}
\label{fig4}
\end{figure}

All except two QSOs were detected in the 24\,$\mu$m band, while
only eight were detected at the 70\,$\mu$m band and one (SDSS
J171550.49+593548.7) at the 160\,$\mu$m band. The low detection
rate of QSOs in far-infrared bands may be because: (1) all QSOs
except one (SDSS J171550.49+593548.7 with redshift 0.066) are far
from us, with redshifts from 0.2 to 2.5; (2) QSOs are quite warm
compared to normal galaxies; (3) the 7$\sigma$ limits of MIPS
catalog (Frayer et al. 2006). The near-to-mid infrared SEDs of
QSOs are plotted in Figure~5 scaled by rest-frame 3.6\,$\mu$m
flux. The large dots with error bars in Figure~5 are the median
values at the corresponding wavelength bins. At wavelengths longer
than 1\,$\mu$m, the SEDs of QSOs are very flat except SDSS
J171550.49+593548.7. In fact, SDSS J171550.49+593548.7 (dotted
line) is a nearby Seyfert 1 galaxy with strong stellar continuum
and PAH emission in mid-infrared. Since we do not have any
representative infrared SEDs of QSOs, we only plot an AGN template
of NGC~5506 (solid line in Fig.\,5) as comparison.

\begin{figure}
\centering
\includegraphics[width=90mm]{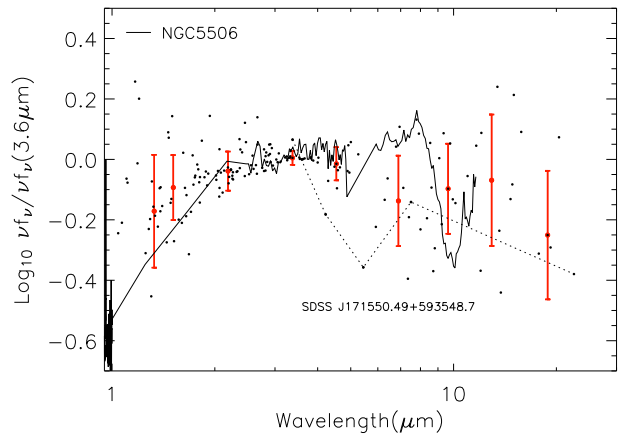}
\vspace{-5mm}

\caption{\baselineskip 3.6mm SEDs of QSOs with fluxes scaled at
rest-frame 3.6\,$\mu$m. The large dots with error bars are median
values and scatter for the wavelength bins. The line is the SED of
the AGN NGC~5506. Most of QSOs are flat and are consistent with
the template of NGC~5506. The dotted line shows the nearest
Seyfert 1 galaxy in our sample.} \label{fig5}
\end{figure}

\section{Results and Discussion}
\label{sect:discussion}
\subsection{Comparison with the Stellar Population Model with Different Age and Fitting Method}

In the SED fitting, we adopted a 13 Gyr underlying stellar
component model from GRASIL (Silva et al. 1998). However, our
sample includes many disk galaxies, which are composed of both a
young population with age about 1 Gyr and an old population.  To
investigate how the age of the stellar population model affects
the fitting results, we compared our 13 Gyr stellar component
model with a 900 Myr model from Vazquez \& Leitherer (2005). We
refitted the SEDs with the dust model plus a 900 Myr stellar
component model, and derived an $\alpha$ values that differed
insignificantly from the previous ones. The mean difference of
$\alpha$ values of two age models is about 0.04 with a scatter of
0.04. Therefore, apparently, the model age does not affect our
fitting.

We have assumed that both the IRAC 3.6\,$\mu$m and 4.5\,$\mu$m
emissions were from the stellar component. As there may exist
3.3\,$\mu$m PAH emission in the IRAC 3.6\,$\mu$m band, it is
necessary to explore the effect of such PAH emission on the
fitting. We changed the fitting by only using 4.5\,$\mu$m flux to
determine the stellar component, instead of using both 3.6\,$\mu$m
and 4.5\,$\mu$m fluxes simultaneously, then refitted the dust
model for all the emission line galaxies. The mean difference of
best fitted $\alpha$ values between the two methods for sample
galaxies is 0.01 with a scatter of 0.03, which is negligible. It
indicates that the 3.3\,$\mu$m PAH emission does not significantly
affect our fitting, either.

\subsection{The $\alpha$ and Mid-To-Far Infrared Properties }

Although Dale et al. (2001) introduced the one parameter dust
model characterized by $\alpha$, they essentially defined the
model by the far-infrared color. The index $\alpha$ characterizes
the distribution of dust mass as a power-law function of local
heating intensity (Dale et al. 2005). It should relate with the
mid-to-far infrared properties of galaxies.

Figure 6 shows the $\alpha$ values of different spectral types as
a function of either mid-infrared or far-infrared dust emission
luminosity. It is clear from all four panels of Figure~6 that
there is no correlation, or just a weak one between $\alpha$ and
the mid- or far-infrared luminosities, for each category of sample
galaxies. However, we can see from Figure~7 that there exist
obvious correlations between $\alpha$ and the ratios of mid- or
far-infrared luminosities to the 3.6\,$\mu$m luminosities.

To quantitatively analyze all these correlations, we applied
Spearman correlation analysis separately to the HII galaxies,
LINERs, mixture types and the whole sample. The results are listed
in Table 2. It is clear from Table 2 that the mid- and
far-infrared luminosities are weakly correlated with $\alpha$ for
the HII galaxies (the null probability being less than 0.06) but
the correlations between $\alpha$ and the ratios of the infrared
luminosities to the 3.6\,$\mu$m luminosity are much stronger. The
probability of no-correlation is less than 0.004 or even smaller.
Furthermore, the correlations are more tight for either the warm
or cold dust continuum at 24\,$\mu$m, 70\,$\mu$m and 160\,$\mu$m
than for the dust emission (dominated by PAH) at 8\,$\mu$m . It
indicates that the parameter $\alpha$ can describe well the shape
of mid-to-far infrared SED. The K band luminosity, is dominated by
old stellar population and has less reddening, we can also adopt
the 3.6\,$\mu$m luminosity as a rough tracer of the stellar mass.
Thus the ratios of mid-to-far infrared luminosities to the
3.6\,$\mu$m luminosity can be regarded as the dust emission
intensity per unit mass. Therefore, the above strong correlations
imply that the $\alpha$ value is strongly related with the dust
emission intensity rather than the total dust emission within the
whole galaxy. One can also see from Table 2 there are no
correlations for the LINERs (all the null probabilities except one
are greater than 0.1).  For the LINERs, the $\alpha$ value does
not depend on either the absolute or relative mid-to-far infrared
dust emission, this implies that LINERs are different from HII
galaxies in the mid-to-far infrared range. As to the mixture
types, the correlation analysis shows that their degrees of
correlation are intermediate between those of HII galaxies and
LINERs. Taken as a whole, the sample galaxies show the same
degrees of correlation as the HII galaxies.

\begin{figure}
\vspace{-5mm} \centering
\includegraphics[width=130mm]{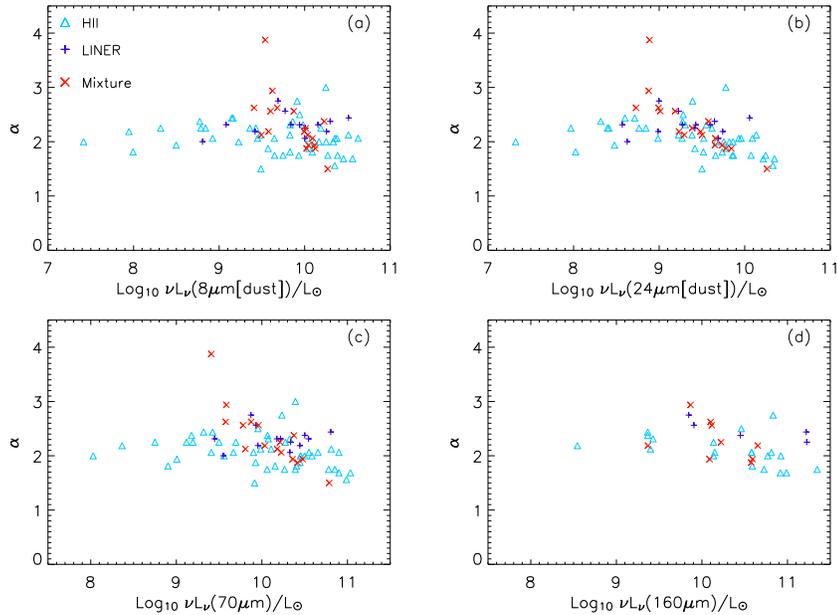}

\vspace{-5mm} \caption{\baselineskip 3.6mm
 Plot of $\alpha$ versus mid- and
far-infrared luminosities: (a) 8\,$\mu$m (dust), (b) 24\,$\mu$m
(dust), (c) 70\,$\mu$m and (d) 160\,$\mu$m. Triangles, plus signs
and crosses represent HII galaxies, LINERs and mixture types,
respectively. } \label{fig6}
\end{figure}

\begin{figure}
\centering
\includegraphics[width=130mm]{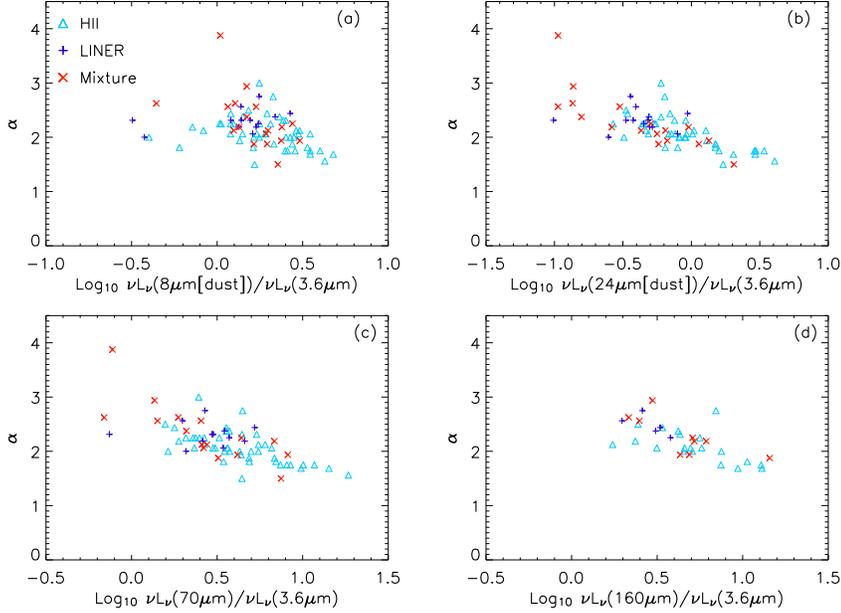}

\vspace{-5mm} \caption{\baselineskip 3.6mm Plot of $\alpha$ values
versus ratios of mid- and far-infrared dust luminosities to the
3.6\,$\mu$m luminosity : (a) 8\,$\mu$m (dust), (b) 24\,$\mu$m
(dust), (c) 70\,$\mu$m and (d) 160\,$\mu$m. The symbols are as
same as in Fig.\,6.} \label{fig7}
\end{figure}

\begin{table}

\centering

\caption[]{The Spearman Correlation Analysis Results of $\alpha$
with Different Parameters for Sample Galaxies}

\fns \tabcolsep 1.5mm
 \begin{tabular}{lccccccccc}
  \hline\noalign{\smallskip}
         &    HII &&                  LINER  &&           Mixture  &&           Whole & \\
            &    $r$ & Prob (null)&    $r$ & Prob (null)&  $r$ & Prob (null)&    $r$ & Prob(null) \\
  \hline\noalign{\smallskip}
$\log~{\nu}L {\nu}
({\rm 8\, \mu m[dust]})/L {\odot}$  & --0.280& 5.7e--02& 0.204&5.0e--01&--0.658&4.1e--03&--0.228& 4.6e--02 \\
$\log~{\nu}L {\nu}
({\rm 24\, \mu m[dust]})/L {\odot}$ & --0.457& 1.3e--03& 0.053&8.6e--01&--0.935&3.9e--08&--0.507& 2.6e--06 \\
$\log~{\nu}L {\nu}
({\rm 70\, \mu m)}/L {\odot}$       & --0.389& 7.6e--03& 0.075&8.1e--01&--0.832&1.2e--04&--0.343& 2.7e--03 \\
$\log~{\nu}L {\nu}
{\rm (160\, \mu m)}/L {\odot}$      & --0.639& 2.4e--03&-0.900&3.7e--02&--0.412&2.7e--01&--0.463& 5.9e--03 \\
 \hline\noalign{\smallskip}
$\log~{\nu}L {\nu}
{\rm (8\, \mu m[dust])}/\nu L {\nu}(\rm 3.6\, \mu m)$ &--0.416&3.7e--03& 0.390&1.9e--01&--0.557&2.0e--02&--0.411&2.0e--04 \\
$\log~{\nu}L {\nu}
{\rm (24\, \mu m[dust])}/\nu L {\nu}(\rm 3.6 \,\mu m)$&--0.682&1.3e--07&--0.100&7.4e--01&--0.856&2.3e--05&--0.702&1.5e--12 \\
$\log~{\nu}L {\nu}
{\rm (70 \,\mu m)}/\nu L {\nu}(\rm 3.6 \,\mu m)$      &--0.657&7.1e--07&--0.045&8.8e--01&--0.817&2.0e--04&--0.640&8.5e--10 \\
$\log~{\nu}L {\nu}
{\rm (160 \,\mu m)}/\nu L {\nu}(\rm 3.6 \,\mu m)$     &--0.689&7.8e--04&--0.800&1.0e--01&--0.689&4.0e--02&--0.692&5.9e--06 \\
  \noalign{\smallskip}\hline\\
  \end{tabular}
{{\bf Note:} $r$ is the correlation coefficient and Prob(null) is the probability that the correlation does not exist.}
\end{table}

\subsection{The $\alpha$ Distributions of Different Spectral Types}

As discussed above, $\alpha$ is a parameter that determines the
shape of the SED in mid-to-far infrared. The different shapes
represent relative contributions of active and quiescent regions
within the galaxy. The coldest galaxies are fitted by $\alpha$=2.5
as estimated for the cirrus-like region, and the warmest galaxies,
by $\alpha$=1 as estimated for photodissociation regions in
star-forming regions (Dale et al. 2001). The $\alpha$ value can be
used to roughly estimate the activity of the galaxy.

Figure 8(a) shows the distribution of $\alpha$ values of  our
sample galaxies (solid line) compared with SINGS galaxies (dotted
line) from Dale et al. (2005). The derived $\alpha$ values of our
sample range from 1.5 to 3.0 except one. Most of the galaxies have
$\alpha$ values above 2 and few galaxies have that less than 1.5.
All these indicate that most of our sample galaxies tend to be
quiescent, similar to the FIRBACK 170\,$\mu$m sources (Dennefeld
et al. 2005), which are moderate starbursters with a dominating
cold dust component. Compared to SINGS galaxies, our sample shows
a lack of the most quiescent galaxies with $\alpha$ above 3. The
$\alpha$ distribution of SINGS galaxies present two distinct
populations: the first peaks at about 2.2 and is similar to our
sample, the second peaks at about 3.7 and has $\alpha$ values even
as high as 4, which belongs to the most quiescent state. The
difference could be that the SEDs of our sample have fewer points
in the far-infrared, and with larger errors, so our weighted
$\chi^2$ fitting leans to the mid-infrared, which have more points
and lower errors, and this would underestimate the $\alpha$ value
of some galaxies, (such a case can be seen in some individual SEDs
with 160\,$\mu$m fluxes above the model values). Another possible
reason is that the redshifts of our sample are larger than those
of SINGS's. With the high limits (7$\sigma$) of MIPS catalog of
xFLS (Frayer et al. 2006), the most quiescent galaxies with larger
redshifts are probably not detected in the far-infrared bands and
this will bias our sample to the more active one.

\begin{figure}[b!!!]

\vspace{-5mm} \centering
\includegraphics[width=120mm]{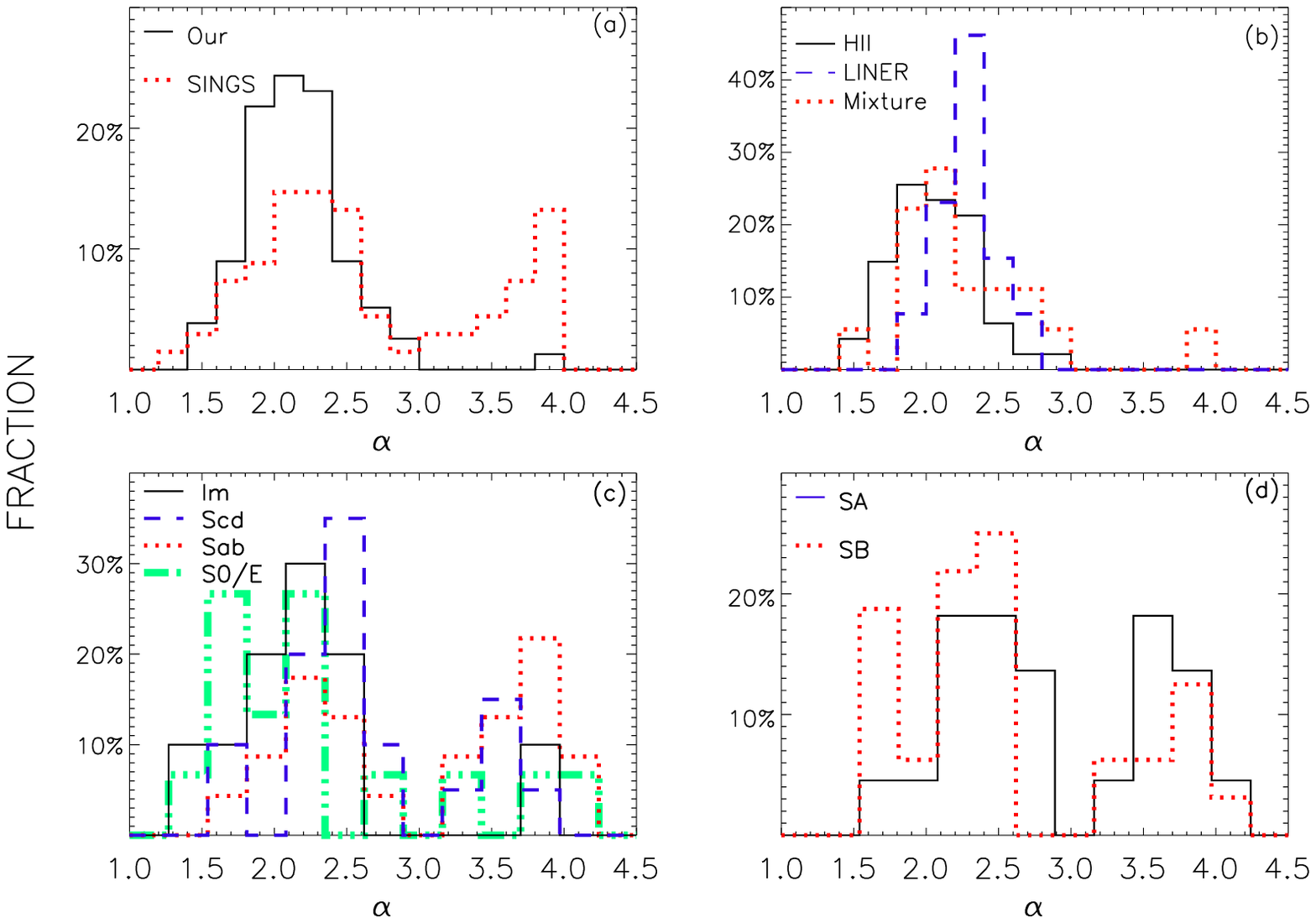}

\caption{\baselineskip 3.6mm  Histograms of $\alpha$ distribution
of different sets of galaxies.  (a) Our whole sample (solid line)
and SINGS sample (dotted line); (b) HII galaxies (solid line),
LINERs (dashed line), and mixture types (dotted line); (c)
galaxies with different Hubble types (solid line for irregular
galaxies, dashed for late type disks, dotted for early type disks,
and dash-dotted for early type galaxies); (d) galaxies with bars
(dotted line) and without bars (solid line).} \label{fig8}
\end{figure}

Figure 8(b) shows the $\alpha$ distributions of HII galaxies
(solid line), LINERs (dashed line) and mixture types (dotted
line). All three types have a similar range of $\alpha$ values,
but differ a little in the mean value: the mean $\alpha$ and
dispersion are ($2.07 \pm 0.29$) for the HII galaxies, ($2.31 \pm
0.19$) for the LINERs and ($2.31 \pm 0.51$) for the mixture types.
One mixture type galaxy (SDSS J172403.46+593804.0) has the largest
$\alpha$ value of 3.88. A K-S test on the $\alpha$ distributions
of LINERs and mixture types show that the probability of these two
distributions coming from same parent distribution is 0.473.
Similar test for the HII galaxies and LINERs gives a probability
of 0.0065. The test shows that the HII galaxies and LINERs have
quite different $\alpha$ distributions, which indicates different
shapes in their mid-to-far infrared SEDs. Based on the $\alpha$
distributions in Figure~8(b)  and the mean $\alpha$ values of the
three spectral types, we find HII galaxies on average are a little
more active than either the mixture types or the LINERs; and the
LINERs are the most quiescent one of the three, and this can also
be supported by the fact that there are very few LINERs with
$\alpha$ value smaller than 2 in our sample, and that the mixture
type as the transition type from HII galaxy to LINER has
properties intermediate between the two. However, though there
exist some differences of different spectral types in mid-to-far
infrared region, we can not firmly distinguish the three types
simply based on the mid-to-far infrared SED in individual cases.

\subsection{The $\alpha$ Distributions of Different Morphological Types}

Since the amount of dust is strongly related to the galaxy's
morphology, the $\alpha$ value would be also expected to relate
with the galaxy morphology. In this section, we explore the
$\alpha$ distributions of galaxies of different morphological
types. Because we lack the morphological information, we can not
do the analysis based on our sample. Fortunately, the SINGS sample
(Dale et al. 2005) provide the morphological data. Figure~8(c)
shows the $\alpha$ distributions of the different morphological
types. All these types exhibit the two populations. According to
the $\alpha$ distributions, the irregular galaxies, on the whole,
are more active (because of their larger star formation density)
than the disk galaxies, and among the latter, the late types are
more active than the early types. In the early type disk galaxies,
the two populations are even comparable in number.  We should
explain that the largest $\alpha$ value would relate to cold
cirrus from the oldest stellar disk. All these results show that
the star formation activity varies along the Hubble sequence. Few
of the early type galaxies (E/S0) show $\alpha$ values greater
than 3. This could be explained by a smaller contribution from
cold cirrus because of the smaller fraction of the disk component
in early type galaxies, but a large, unbiased sample and a
quantitative morphological classification are needed to check
these points.

We also explored the influence of bars on the $\alpha$
distributions (Fig.\,8(d)). We found that bar galaxies tend to
have smaller $\alpha$ values and are a little more active than the
non-bar galaxies. It indicates that bars could trigger off star
formation in galaxies.

\section{Conclusions}
\label{sect:conclusion}

We present
mid-to-far infrared SEDs of a large sample of
galaxies,
selected from the Spitzer xFLS field and SDSS spectroscopic survey. With the help of model fitting, we analyze the dust properties of galaxies in different
spectral types statistically.

The
QSOs present
a power-law continuum at near-to-mid infrared, which differs from SEDs of other galaxies and is well consistent with an AGN template of NGC~5506.

The mid-infrared SEDs of absorption galaxies are dominated by
an old stellar population with little dust emission, and can be well
fitted by the spectrum of
the 13 Gyr dust-free elliptical galaxy.

The mid-to-far infrared SEDs of HII galaxies, mixture type
galaxies, and LINERs can be well fitted by
the one-parameter dust
model of Dale et al. (2001)
plus
the 13 Gyr dust-free elliptical galaxy model. The statistics shows
that
the HII galaxies
are more active than the other two types, though most of these galaxies should be classified as quiescent galaxies.

The $\alpha$ distributions of different morphologies of SINGS
galaxies show that the activity of galaxies
depends
on
the
Hubble morphological
type
and the presence of bars.

\begin{acknowledgements}
The authors thank
for the help from Cai-Na Hao, Jian-Ling Wang and
Feng-Shan Liu. Many thanks go to D.A. Dale for kindly providing
the dust model. Also many thanks go to the referee, who gives
suggestions to improve the paper. This project was supported by
the NSFC through Grants 10273012, 10333060 and 10473013. This work
is based in part on observations made with the {\it Spitzer Space
Telescope}, which is operated by the Jet Propulsion Laboratory,
California Institute of Technology under NASA contract 1407. The
SDSS Web site is http://www.sdss.org/. SDSS is managed by the
Astrophysical Research Consortium (ARC) for the Participating
Institutions.
\end{acknowledgements}


\begin{thebibliography}{99}
\small
\setlength{\itemindent}{-3mm}
\setlength{\itemsep}{-0.5mm} 
\setlength{\baselineskip}{4.5mm}

\bibitem[]{} Bertin E., Arnouts S., 1996, A\&AS, 117, 393
\bibitem[]{} Calzetti D., Armus L., Bohlin R. C. et al., 2000, ApJ, 533, 682
\bibitem[]{} Calzetti D., 2001, PASP, 113, 1449
\bibitem[]{} Dale D. A., Helou G., Contursi A. et al., 2001, ApJ, 549, 215
\bibitem[]{} Dale D. A., Helou G., 2002, ApJ, 576, 159
\bibitem[]{} Dale D. A., Bendo G. J., Engelbracht C. W. et al., 2005, ApJ, 633, 857
\bibitem[]{} Dennefeld M., Lagache G., Mei S. et al., 2005, A\&A, 440, 5
\bibitem[]{} Fazio G. G., Hora J. L., Allen L. E. et al., 2004, ApJS, 154, 10
\bibitem[]{} Frayer D. T., Fadda D., Yan L. et al., 2006, AJ, 131, 250
\bibitem[]{} Hatziminaoglou E., P\'erez-Fournon I., Polletta M. et al., 2005, AJ, 129, 1198
\bibitem[]{} Huang J. S., Barmby P., Fazio G. G. et al., 2004, ApJS, 154, 44
\bibitem[]{} Kauffmann G., Heckman T. M., Tremonti C. et al., 2003, MNRAS, 346, 1055
\bibitem[]{} Kennicutt R. C. J., Armus L., Bendo G. et al., 2003, PASP, 115, 928
\bibitem[]{} Lacy M., Wilson G., Masci F. et al., 2005, ApJS, 161, 41
\bibitem[]{} Lagache G., Dole H., Puget J. L., 2003, MNRAS, 338, 555
\bibitem[]{} Lonsdale C. J., Smith H. E., Rowan-Robinson M. et al., 2003, PASP, 115, 897
\bibitem[]{} Oke J. B., Gunn J. E., 1983, ApJ, 266, 713
\bibitem[]{} Rieke G. H., Young E. T., Engelbracht C. W. et al., 2004, ApJS, 154, 25
\bibitem[]{} Rowan-Robinson M., Babbedge T., Surace J. et al., 2005, AJ, 129, 1183
\bibitem[]{} Rowan-Robinson M., Lari C., Perez-Fournon I. et al., 2004, MNRAS, 351, 1290
\bibitem[]{} Sajina A., Scott D., Dennefeld M. et al., 2006, MNRAS, 369, 939
\bibitem[]{} Silva L., Granato G. L., Bressan A. et al., 1998, ApJ, 509, 103
\bibitem[]{} Soifer B. T., Neugebauer G., Helou G. et al., 1984, ApJ, 283, L1
\bibitem[]{} Stoughton C., Lupton R. H., Bernardi M. et al., 2002, AJ, 123, 485
\bibitem[]{} Tremonti C. A., Heckman T. M., Kauffmann G. et al., 2004, ApJ, 613, 898
\bibitem[]{} Vazquez G. A., Leitherer C., 2005, ApJ, 621, 695
\bibitem[]{} Veilleux S., Osterbrock D. E., 1987, ApJS, 63, 295
\bibitem[]{} Werner M. W., Roellig T. L., Low F. J. et al., 2004, ApJS, 154, 1
\bibitem[]{} Wu H., Cao C., Hao C. N., et al., 2005, ApJ, 632, L79
\bibitem[]{} Wu H., Zou Z. L., Xia X. Y. et al., 1998, A\&AS, 132, 181
\bibitem[]{} Xilouris E. M., Madden S. C., Galliano F. et al., 2004, A\&A, 416, 41

\end{thebibliography}
\end{document}